\documentstyle[12pt,fleqn]{article}
\input epsf
\def\singlespace {\smallskipamount=3.75pt plus1pt minus1pt
                  \medskipamount=7.5pt plus2pt minus2pt
                  \bigskipamount=15pt plus4pt minus4pt
                  \normalbaselineskip=15pt plus0pt minus0pt
                  \normallineskip=1pt
                  \normallineskiplimit=0pt
                  \jot=3.75pt
                  {\def\smallskip {\vskip\smallskipamount}}
                  {\def\medskip   {\vskip\medskipamount}}
                  {\def\bigskip   {\vskip\bigskipamount}}
                  {\setbox\strutbox=\hbox{\vrule 
                    height10.5pt depth4.5pt width 0pt}}
                  \parskip 7.5pt
                  \normalbaselines}
\def\middlespace {\smallskipamount=5.625pt plus1.5pt minus1.5pt
                  \medskipamount=11.25pt plus3pt minus3pt
                  \bigskipamount=22.5pt plus6pt minus6pt
                  \normalbaselineskip=22.5pt plus0pt minus0pt
                  \normallineskip=1pt
                  \normallineskiplimit=0pt
                  \jot=5.625pt
                  {\def\smallskip {\vskip\smallskipamount}}
                  {\def\medskip   {\vskip\medskipamount}}
                  {\def\bigskip   {\vskip\bigskipamount}}
                  {\setbox\strutbox=\hbox{\vrule 
                    height15.75pt depth6.75pt width 0pt}}
                  \parskip 11.25pt
                  \normalbaselines}
\def\doublespace {\smallskipamount=7.5pt plus2pt minus2pt
                  \medskipamount=15pt plus4pt minus4pt
                  \bigskipamount=30pt plus8pt minus8pt
                  \normalbaselineskip=30pt plus0pt minus0pt
                  \normallineskip=2pt
                  \normallineskiplimit=0pt
                  \jot=7.5pt
                  {\def\smallskip {\vskip\smallskipamount}}
                  {\def\medskip   {\vskip\medskipamount}}
                  {\def\bigskip   {\vskip\bigskipamount}}
                  {\setbox\strutbox=\hbox{\vrule 
                    height21.0pt depth9.0pt width 0pt}}
                  \parskip 15.0pt
                  \normalbaselines}

\include{dspace12}
\def\be{\begin{equation}}
\def\ee{\end{equation}}
\def\bea{\begin{eqnarray}}
\def\eea{\end{eqnarray}}

\def\sect #1{\setcounter{equation}{0}}

\begin{document}
\singlespace
\rightline{\bf TIFR-TAP Preprint}

\begin{center}
{\Large {The Structure of Singularity in 
Spherical Inhomogeneous Dust Collapse}}
\end{center}
\vspace{1.0in}
\vspace{12pt}

\begin{center}
{\large{ S. Jhingan\footnote{E-mail : sanju@tifrvax.tifr.res.in} and P. S.
Joshi\footnote{E-mail : psj@tifrvax.tifr.res.in}\\
Theoretical Astrophysics Group\\
Tata Institute of Fundamental Research\\
Homi Bhabha Road, Colaba, Bombay 400005, India.}}
\end{center}

\vspace{1.3in}
%%%%%%%%%%%%%%%%%%%%%%%%%%%%%%%%%%%%%%%%%%%%%%%%%%%%%%%%%%%%%%%
\begin{abstract}

We study here the structure of singularity forming in  
gravitational collapse of spherically symmetric inhomogeneous dust.
Such a collapse is described by the Tolman-Bondi-Lema{\^i}tre metric, which 
is a two-parameter family of solutions to Einstein equations, 
characterized by two free functions of the radial coordinate, namely the 
`mass function' $F(r)$ and the `energy function' $f(r)$.  The main new result
here relates, in a general way, the formation of black holes and naked
shell-focusing singularities resulting as the final fate of 
such a collapse to the generic form of regular initial data. Such a data is
characterized in terms of the density and velocity profiles of the matter, 
specified on an initial time slice from which the collapse commences.
Several issues regarding the strength and stability of these singularities, 
when they are naked, are examined with the help of the analysis developed here.
In particular, it is seen that strong curvature naked singularities can
develop from a generic form of initial data in terms of the initial
density profiles for the collapsing configuration. We also establish here
that similar results hold for black hole formation as well. We also discuss
here the physical constraints on the initial data for avoiding shell-crossing 
singularities; and also the shell-focusing naked singularities, so that the
collapse will necessarily end as a black hole, preserving the cosmic 
censorship. These results generalize several earlier works on 
inhomogeneous dust collapse as special cases, and provide a clearer
insight into the phenomena of black hole and naked singularity
formation in gravitational collapse.
 
\end{abstract}
%%%%%%%%%%%%%%%%%%%%%%%%%%%%%%%%%%%%%
\newpage
%%%%%%%%%%%%%%%%%%%%%%%%%%%%%%%%%%%%%
\doublespace
%%%%%%%%%%%%%%%%%%%%%%%%%%%%%%%%%%%%%

\centerline{\bf I. Introduction}

The classic paper by Oppenheimer and Snyder \cite{OSD} analyzes
the gravitational collapse of a spherically symmetric massive object 
to conclude that the collapse of a homogeneous dust cloud would result
into a black hole in the spacetime. Such a black hole is characterized
by the presence of an event horizon, and the spacetime singularity at
the center, which is covered by the horizon. This scenario provides the basic
motivation for the black hole physics, and the cosmic censorship conjecture
\cite{Pen69}, which states that even when the assumptions contained in the
above case are relaxed, in the form of either perturbations in the 
symmetry or form of matter etc., the outcome would still be a black hole in
generic situations. As no proof, or a rigorous mathematical formulation
of the censorship hypothesis has been available so far, 
the course of action that
has emerged in past decade or so has been to carry out a detailed
investigation of different gravitational collapse scenarios in general
relativity to understand the final fate of collapse ( see e.g. \cite {Global}
and references therein).

Within such a context, an important situation that immediately offers 
itself for analysis is the introduction of inhomogeneities in the matter
distribution. It is clear that the assumption of homogeneity is only an
idealization, and realistic density profiles, for massive objects such as
stars will have inhomogeneous density distribution, peaked typically at
the center of the object. One would like to examine the final fate of
gravitational collapse of such an inhomogeneous dust cloud, 
and examine in what ways there are
departures in conclusions as opposed to the homogeneous situation. The 
Einstein equations can be fully solved for the stress-energy tensor of the 
form of inhomogeneous dust, and such a collapse is described by the 
Tolman-Bondi-Lema{\^i}tre (TBL) metric. This is a two-parameter family of
solutions, 
characterized by two free functions of the radial coordinate which are the 
`mass function' $F(r)$ and the `energy function' $f(r)$ of the cloud. While
the first quantity describes here the initial density distribution for
the collapsing cloud, the second function characterizes the initial
velocity profile of the collapsing shells.   

Our main purpose here is to relate, in a general way, the formation 
of black holes and naked shell-focusing singularities resulting as the 
final fate of 
such a collapse to the generic form of regular initial data. Such a data is
characterized in terms of the density and velocity profiles of the matter,
as given by the two functions above, specified on an initial time slice
from which the collapse commences.
An important issue, when naked singularities do form in gravitational
collapse, is that of their curvature strength. 
We discuss here several issues regarding the strength 
and stability of such singularities, 
with the help of the analysis developed here.
In particular, it is seen that strong curvature naked singularities can
develop from a generic form of initial data in terms of the initial
density profiles for the collapsing configuration. We also establish 
here that similar results hold
for black hole formation as well. Thus, we show that 
both the black hole as well as
naked singularity formation are related to the nature of the regular
initial data defined at the onset of gravitational collapse. We also 
discuss here the physical constraints on initial data for avoiding 
shell-crossing singularities; and also the shell-focusing naked singularities,
so that the collapse will necessarily end as a black hole in conformity
with the cosmic censorship conjecture. 
These results generalize several earlier works on 
inhomogeneous dust collapse as special cases, and provide a clearer
insight into the phenomena of the black hole and naked singularity
formation in gravitational collapse. For example, while some of the earlier
works ( see e.g. \cite {Chris, ES, Newm, Josh93}) discuss the formation of
naked singularities in dust collapse, either under various special 
assumptions or under general conditions, they do not address the important
physical problem of relating this naked singularity formation to the
nature of regular initial data at the onset of the gravitational collapse.
It is crucial, from the physical point of view, to characterize and
identify the properties of initial data which distinguish between the
formation of black holes and the occurrence of naked singularity. Such an
analysis could also pave the way for any possible formulation and proof
for the cosmic censorship conjecture. A beginning in such a direction was
made in \cite {JTP, Singh}. We present here a more general and complete
analysis, obtaining several new results in connection with the naked
singularity and black hole formation in gravitational collapse, in the
specific context of the genericity of the initial data.

The main issue of concern here to us is, given a generic density 
distribution at the onset of collapse what
{\it all} is possible as the final outcome. It is seen that given such a
generic density profile, one can always have either a black hole or
a strong curvature naked singularity, depending on the nature of the 
initial data. Certain points related to genericity are also discussed.

Section II below provides a brief review of the TBL models,
giving the main set of equations used. Section III specifies the generic
initial data to be used for collapse, and discusses the formation of 
singularities in collapse situations. Shell-crossing singularities are 
discussed in Section IV. The formation of black holes and naked singularities
is discussed in Section V, and the last section gives certain concluding 
remarks on our results. 
\bigskip

\centerline{\bf II. Tolman-Bondi-Lema{\^i}tre Dust Models}

\bigskip

The TBL spacetimes are spherically symmetric manifolds
($\cal M$,g), with metric of the form,

\be
ds^2 = -dt^2 + e^{2\omega} dr^2 + R^2 d\Omega^2,
\ee

and energy-momentum tensor of the form of a perfect fluid with
equation of state p=0, given by 

\be
{T^{ij} = \epsilon \,\delta^{i}_{t}\delta^{j}_{t}}.
\ee

Here $\epsilon$, $\omega$ and R are functions of  $r$ and  $t$, and
$d\Omega^2$ is the metric on the 2-sphere. The Einstein
equations become

\be
{\dot R}^2 = \frac{F(r)}{R}  + f(r), 
\ee

\be
e^{2\omega} \, (1 + f(r)) = R'^2,
\ee

\be
\epsilon(r,t)= \frac{F'(r)}{R^2 R'},
\ee

where the dot and prime signify partial derivatives with respect to
$t$ and $r$ respectively. 
It is seen that the energy density blows up either at the ``shell-focusing
singularity'' $R=0$, or when $R'=0$ which corresponds to a ``shell-crossing
singularity'' in the spacetime.
The functions $F(r)$ and $f(r)$ are free
functions of integration which completely specify the
initial data in the model. Here $F(r)$ can be interpreted as the weighted mass
given by

\be
F(r) = \int_{{\cal B}_{r,t}}(1+f) \epsilon(r,t) d{\upsilon}_t = 4 \pi
\int_{0}^{r} \rho r^2 dr
\ee

where $\rho = \epsilon(0,r)$, and ${\cal B}_{r,t}$ is a ball of coordinate
radius r centered on r=0 in the hypersurface $\Sigma_{t}$($t$ = const.). 
The time slicing ($t$=const.) has been so chosen such that
$r$=const. labels a matter shell with $t$ measuring the proper
time elapsed along its geodesic path, that is we write the metric in
comoving coordinates. Equation (3) can be thought of as the general
relativistic generalization of the Newtonian energy equation \cite {Bondi},
which leads to the interpretation of $F(r)$ as the effective mass contained 
in the sphere of radius $r$; and $f(r)$ as effective
total energy per unit mass of a fluid element labeled by the
comoving radial coordinate $r$. The model is said to be bound,
unbound, or marginally bound if $f(r)$ is less than zero, greater than
zero, or equal to zero respectively. 

The integrated form of equation (3)  is given by

\be
t - t_s(r) = -\frac{R^{3/2} G(-fR/F)}{\sqrt F}
\ee

where $G(y)$  is a real, positive and smooth function which is bounded,
monotonically increasing and strictly convex, and is  given by

\be   
G(y) =    \cases { {{\rm arcsin}\sqrt{y}\over y^{3/2}} - 
                     {\sqrt{1-y}\over y}, &  $1\geq y> 0$,\cr
                     {2\over 3}, & $y=0$,\cr
                     {-{\rm arcsinh}\sqrt{-y}\over (-y)^{3/2}} - 
                     {\sqrt{1-y}\over y}, & $0> y\geq -\infty$.\cr} 
\ee

and $t_s(r)$ is a constant of integration which can be fixed by the
choice of scaling on the initial surface ($t=0$). Using this scaling freedom,
if we choose so that $ R(0,r)= r$ then,

\be
t_s(r) = \frac{r^{3/2} G(-fr/F)}{\sqrt F}
\ee

The time $t=t_s(r)$ corresponds to the value $R=0$ 
where the area of the matter
shell at a constant value of the coordinate $r$ vanishes, which
corresponds to the physical spacetime singularity. Thus the range of
coordinates is given by

\be
0 \, \leq \, r \, < \, r_c, \, -\infty \, \leq \, t \, < \,
t_{s}(r),
\ee
where $r=r_c$ denotes the boundary of the dust cloud where the 
solution is matched to the exterior Schwarzschild solution.

The argument of the function $G(-fR/F)$ changes it's sign depending on the
sign of $f(r)$, since both $R$ and $F(r)$ take only non-negative values. If
$f(r) \geq 0$ for all $r \geq 0$, then from equation (3) 
we get ${\dot R} \neq 0$ for
all $t$. From the integrated form of equation (3) we can show that if for
some $r > 0$ we have $f(r) > 0$ then this particular shell is strictly
unbounded, and from a physical point of view, the collapse 
can be interpreted as not being gravitational, but due to specific
initial conditions (infinitely far in the past) \cite {Newm}. Therefore,
we shall discuss here only the more physical case with 
$f(r) < 0$ (however, it will be
possible to generalize the conclusions here to other cases as well by
similar methods as used here). 

It is seen that if the condition 

\be
R' > 0
\ee
is satisfied then the dust collapse always avoids shell-crossing 
singularities. The restrictions implied by this condition on the initial 
data will be
discussed in sections IV and V. The weak energy condition, i.e.
$T_{ij} = \epsilon(r,t)V^i V^j \geq 0$, for all nonspacelike vectors,
$V^i$ is assumed everywhere in spacetime. This imposes a condition on
the initial data that the energy density $\epsilon (r,t)$ is non-negative
everywhere. 

\bigskip

\centerline{\bf III. Generic Initial Data and Formation of Singularities}
\bigskip

We call a spacetime to be singular \cite { Gero, Clrkbook} if it
contains an incomplete nonspacelike geodesic,

$$
\gamma : [0,x) \rightarrow {\cal M}
$$ 

(where $\cal M$ is the spacetime manifold) such that there is no extension  

$$
\theta : {\cal M} \rightarrow {\cal M}'
$$

for which $\theta \circ \gamma$ is extendible. The spacetime should
be inextendible,  that is, we should not be able to extend the above
incomplete curve by embedding the spacetime into a larger manifold.
This essentially means that we do not admit singularities created by
removing pieces from the spacetime manifold.
Important theorems showing the existence of singularities
in a spacetime under ``reasonable" physical conditions, which might form
either in gravitational collapse or cosmology, were proved by Penrose, 
Hawking, and Geroch \cite {Haw, 
HawEll}. While these theorems prove the existence of singularities under
a fairly broad spacetime framework, they provide no information on
the nature and structure of the singularities they predict, either in collapse
scenarios or cosmology. 
In particular, they do not imply whether the singularities
resulting as the final fate of gravitational collapse will be necessarily 
covered, or otherwise, by the event horizons of gravity. 
Our purpose here is to analyse the nature and structure of curvature
singularities occurring in the TBL collapse models. A singularity
will be  called naked if there are future directed
nonspacelike curves in the spacetime which terminate at the 
singularity in the past, otherwise it will be hidden behind  a black hole. 
Considering equation (7),
the apparent horizon within the collapsing dust cloud lies at $R=F(r)$, i.e.

\be
t_{ah}(r) = t_{s}(r) - FG(-f)
\ee

It can be easily seen from the above equation that all the points on the
singularity curve $t_s(r)$, other than the central point ($r=0$) are
covered by the apparent horizon. This is because, since both the functions 
$F(r)$ and $G(r)$ are strictly positive for $r > 0$, with
$F(r)=0$ at $r=0$, therefore for all $ r > 0 $ $\;$ $ t_{s}(r) >
t_{ah}(r) $ and $t_{s}(0) = t_{ah}(0)$. Thus, all the other points on 
the singularity curve, except the point $r=0$, are covered by the 
apparent horizon. It is the central
singularity point $r = 0$ whose nature, in terms of being naked or
covered, depends on the data specified at the initial surface.

In the TBL models the initial data, to be specified on the initial
hypersurface $\Sigma_i$, consists of two free functions $F(r)$ and $f(r)$
respectively and such a specification 
uniquely determines the solution to field equations
inside  $D^+(\Sigma_i)$, which is the future Cauchy development of the 
hypersurface $\Sigma_i$.  To study the formation of singularity for this
initial data, we consider the initial functions in a
``generic" (most general expandable) form
 
\be
F(r) = \sum_{n=0}^{\infty}F_n r^{n+3}, \: \: f(r) = \sum_{n=2}^{\infty}f_n
r^{n}.
\ee

The choice of $n$ here is made to avoid any singular behaviour of functions,
and to ensure the finiteness of density,
on the initial surface. The functions chosen here are assumed to be
$C^2$ only, and not
necessarily analytic ($C^{\omega}$); since the only requirement of the theory
is to have $C^2$ functions rather than restricting to the stronger
differentiability condition of analyticity. 
This way, we are able to deal with a broader
class of spacetimes. Additional differentiability requirements higher 
than $C^2$  will make the situation functionally less generic. 
We now recall briefly some definitions \cite {Josh93} which are necessary 
for further analysis. Especially, the partial derivative
for the area function $R$ is written as,

\be
\begin{array}{clcr}
R'= &r^{\alpha - 1}[(\eta - \beta) X + [\Theta - (\eta - \frac{3}{2}\beta)
X^{3/2} G(-PX)]\times[P + \frac{1}{X}]^{1/2}]\\
 \equiv &r^{\alpha - 1}H(X,r) 
\end{array}
\ee

where we have used the following

%\be
%\begin{array}{clcr}
%u = r^{\alpha}, \; X=(R/u),& \eta(r) = r\frac{F'}{F}, \; \beta(r) =
%r\frac{f'}{f}, \; p(r) = r\frac{f}{F}, \; P(r)=p r^{\alpha-1},\\  
%\Lambda = \frac{F}{r^{\alpha}},& \Theta \equiv \frac{t'_s \sqrt
%\Lambda}{r^{\alpha-1}} =
%\frac{1+\beta-\eta}{(1+p)^{1/2}r^{3(\alpha-1)/ 2}} +
%\frac{(\eta - \frac{3}{2}\beta)G(-p)}{r^{3(\alpha-1)/ 2}} 
%\end{array}   
%\ee  

$$
u = r^{\alpha}, \; X=(R/u),  \; \eta(r) = r\frac{F'}{F}, \; \; \beta(r) =
r\frac{f'}{f}, \; \; p(r) = r\frac{f}{F}, \; \; P(r)=p r^{\alpha-1}, \; \; 
$$

\be
~~~\Lambda = \frac{F}{r^{\alpha}}, \; \; \; \Theta \equiv \frac{t'_s \sqrt
\Lambda}{r^{\alpha-1}} =
\frac{1+\beta-\eta}{(1+p)^{1/2}r^{3(\alpha-1)/ 2}} +
\frac{(\eta - \frac{3}{2}\beta)G(-p)}{r^{3(\alpha-1)/ 2}} 
\ee

The function $\beta(r)$ is defined to be zero when $f(r)$ is zero 
(the marginally
bound case). The parameter $\alpha$ (which satisfies  $\alpha\geq 1$)
is introduced here for examining
the structure of the central  singularity at $r=0$. In terms of these new
variables, we can write for energy density as 

\be
\epsilon=\frac{\eta \Lambda}{R^2 H}
\ee

The weak energy condition implies $H(X,r) \geq 0$, and $\eta \Lambda \geq
0$. The actual value of $\alpha$ is uniquely determined by the initial
data prescribed, and is the key factor in deciding whether or not we have
a strong curvature (covered or naked) singularity.
The expression for energy density on the initial surface (which is the scaling
surface $\Sigma_i$, on which $R=r$) is $\epsilon = F'/ r^2$, and 
the weak energy condition
implies  that $F'\geq 0$. The requirement that initial surface
should not contain any trapped surfaces $(F(r)/R(r,t) > 1)$
also gives a restriction on the choice of initial data.

The equation for radial null geodesics in the spacetime (1) is given by

\be
\frac{dt}{dr} = \pm \frac{R'}{(1 + f(r))^{1/2}}
\ee

This can also be written in the form,

\be
\frac{dR}{du} = (1 - \frac{\sqrt{f + \Lambda/X}}{\sqrt{1 + f}})
\frac{H(X,r)}{\alpha} \equiv U(X,r)
\ee

Defining

\be
X_0 = \; \lim_{\stackrel{R \rightarrow 0}{u \rightarrow 0}} \frac{R}{u}
\; = \;\lim_{\stackrel{R \rightarrow 0}{u \rightarrow 0}} \frac{dR}{du}
\ee

it can be shown \cite {Josh93} that if the equation

\be
V(X) = U(X,0) - X = (1 - \frac{\sqrt{f_0 + \Lambda_0/X}}{\sqrt{1 + f_0}})
\frac{H(X,0)}{\alpha} - X = 0
\ee

admits a real positive root, then the central singularity at
$r=0, R=0$ is naked (at least
locally). The global visibility of such a singularity will depend on the
overall behaviour of the concerned functions within the dust cloud in the
range $0<r<r_c$. In the case otherwise, a black hole will be formed as
the end product of collapse. The parameter $\alpha$ is
uniquely fixed by demanding that $\Theta(X,r)$  goes to a nonzero finite
value in the limit $r \rightarrow 0$. Consider the expression for $\Theta$

\be
\Theta \equiv \frac{t'_s \sqrt \Lambda}{r^{\alpha-1}} =
\frac{1+\beta-\eta}{(1+p)^{1/2}r^{3(\alpha-1)/ 2}} +
\frac{(\eta - \frac{3}{2}\beta)G(-p)}{r^{3(\alpha-1)/ 2}} =
\frac{\Psi(r)}{r^{3(\alpha-1)/2}}
\ee
where

\be
\Psi(r) = \Psi_0 + \Psi_1 r + \Psi_2 r^2 + \Psi_3 r^3 + \Psi_4 r^4 + \cdot
\cdot \cdot \cdot
\ee

The value of $\alpha$ is decided by the first nonvanishing term in the
expansion of $\Psi(r)$. Each term in the expansion of $\Psi(r)$ is
completely specified in terms of the initial data; so given such data at the
initial surface in terms of the density and velocity distributions,
as specified by the functions $F$ and $f$ above, 
we can tell using the criterion above whether the final fate of
collapse is going to be a  black hole or a naked
singularity.

The nonspacelike geodesics of the spacetime (1) are given by, 

\be
K^t = \frac{dt}{dk} = \frac{\cal P}{R}
\ee

\be
K^r = \frac{dr}{dk} = \frac{\sqrt{1+f}\sqrt{{\cal P}^2 - l^2 + B
R^2}}{RR'}
\ee

\be
(K^{\theta})^2 + sin^2{\theta}(K^{\phi})^2 = l^2/R^4
\ee

where the $K^i = dx^i/dk$ denote the tangent to the geodesics. 
Here $k$ is affine parameter along geodesics, $l$ is the impact parameter
($l=0 \Rightarrow$ radial trajectories), and the values 
$B=0,-1$ correspond to null and timelike curves
respectively. The function $\cal P$ satisfies the differential equation

\be
\frac{d{\cal P}}{dk} + ({\cal P}^2 - l^2 + BR^2)[\frac{{\dot R'}}{RR'} -
\frac{{\dot R}}{R^2}] - ({\cal P}^2 - l^2 + BR^2)^{1/2}{\cal
P}\frac{\sqrt{1+f}}{R^2} + B{\dot R} = 0
\ee

For the clarity of discussion, we consider only radial null geodesics 
here, in which case the above equation reduces to

\be
\frac{1}{{\cal P}^2}\frac{d{\cal P}}{dk} + [\frac{{\dot R'}}{RR'} -
\frac{{\dot R}}{R^2}] - \frac{\sqrt{1+f}}{R^2} = 0
\ee

We shall call the singularities to be {\it strong} (see e.g. {\cite
{Clrkbook, Tip77}) if along nonspacelike trajectories the following is
satisfied,

\be
\lim_{k \rightarrow 0} k^2 \psi(r) = \lim_{k \rightarrow 0} k^2 R_{ij} V^i
V^j \neq 0 
\ee

Such a rate of divergence 
indicates a very powerful curvature growth in the limit of approach
to the singularity (which is the same as in the case of the big bang
singularity of cosmology), 
and ensures that all the volume forms, as defined by the 
Jacobi fields along the nonspacelike geodesics, are crushed to zero size
in the limit of approach to the singularity. Physically, this can be 
interpreted as indicating that all the objects falling into the singularity
are crushed to zero size. The significance of this would be that there
could not possibly be any extension of the spacetime through such a
singularity, as opposed to a gravitationally weak singularity such as a
shell-cross through which the spacetime could possibly be extended and
continued further.

In case of TBL models, we have 

\be
{\lim_{k \rightarrow 0}} k^2 \psi(r) = {\lim_{k \rightarrow 0}}
\frac{F'({K^t})^2}{R^2R'} 
\ee

For radial null geodesics, we can write using the expressions above
for $K^t$, $R'$, and using the L'Hospital rule,

\be
\lim_{k \rightarrow 0} k^2 \psi(r)=
\lim_{k \rightarrow 0} \frac{4 \eta_0 \Lambda_0 H_0}{{X_0}^2[\sqrt{1+f_0}
(3\alpha - \eta_0)- N_0]^2}
\ee
Where the quantity $N_0$ is the limiting value of $-r\dot R'$, and 
$\Lambda_0$ is given by

\be   
\Lambda_0 =    \cases { 0,      & $\alpha < 3$,\cr 
                      {F_0},    & $\alpha = 3$,\cr
                      {\infty}, & $\alpha > 3$.\cr}
\ee

Since $\Lambda_0$ occurs in the numerator of expression for $k^2 \psi(r)$,
we  have strong curvature singularities only for $\alpha
\geq 3$. All the same, it follows from equation (18) that whenever 
$\alpha$ is greater than 3 we always have black hole, since we cannot have any
outgoing geodesics meeting the singularity in the past in that case.
\bigskip

\centerline{\bf IV. Shell-Crossing Singularities}
\bigskip

Some counterexamples to cosmic censorship were proposed \cite {YodSze} 
using  the so called shell-crossing singularities by Yodzis et. al., who
showed the existence of such singularities and  also that they are naked. 
However, these singularities are 
gravitationally weak (both in Tipler \cite {Tip77} as well as the 
Kr{\'o}lak \cite {Krol} sense); and there 
have been proposals for extending the spacetime through such singularities, 
in particular, by Papapetrou and Hamui \cite {Papa} and also others. 
These singularities 
are generally not considered as being any serious counterexample to 
cosmic censorship conjecture, as the spacetime may be extended in a 
distributional sense through such a mild singularity. However, it
is of importance to see what are the restrictions imposed on the initial data 
and the spacetime by the avoidance of such singularities at any point 
other then at the
center, before the central shell-focusing singularity occurs; and
to examine how `physical' or `unphysical'
such restrictions are. That is, we would like to know what constraints 
the avoidance these singularities places on the functional form of 
initial data.  We will impose the criterion
for avoidance of shell-cross singularities on the initial data for making
it physically more reasonable.

Shell-crossings are singularities where we observe crossing of
dust shells, and the density diverges at these points. Consider the
expression for energy density (5) in the TBL models.
We have a shell-crossing singularity at $R'=0$ and at this point 
there is a divergence in density and also certain 
curvatures. But for the spherically symmetric dust collapse 
it is possible to find a
coordinate system so that we have a regular $C^1$ extension of the 
metric through these singularities,
but which need not be $C^2$ \cite {Newm, Lun}. 
Calculating the expression for $R'$ using equation (7),

\begin{eqnarray}
[\frac{3 q}{2} (\frac{R}{r})^{1/2}G(sR/r) + {q s}
(\frac{R}{r})^{3/2}G^{1}(sR/r)]R'& = & r q s'[G^1(s) - (\frac{R}{r})^{5/2}
G^1(sR/r)] \nonumber \\  + q (\frac{R}{r})^{3/2}[\frac{3 G(sR/r)}{2} +
p(\frac{R}{r})G^1(sR/r)] & - & 
\frac{r q'}{2}[G(s) - (\frac{R}{r})^{3/2} G(sR/r)]
\end{eqnarray}

where we have used notation, 

\be
G^1(x) = dG(x)/dx, \; \; s(r)=-\frac{r f(r)}{F(r)}, \; \;
q(r)=\frac{F(r)}{r^3}
\ee

Clearly, if $G(x)$ is strictly positive and convex, and  $R/r \leq 1$ for
$t>0$,  we cannot have shell crossings \cite {Newm}  if

\be
s' \geq 0 \; , \; q' \leq 0
\ee

The restrictions imposed by the above equations are physically reasonable
in the following sense. The condition $q' \leq 0$ 
implies that the density should be constant or decreasing away
from the center, which gives a restriction on the nonvanishing terms in
the expansion for density. Now analysing
the condition $s' \ge 0$, the kinetic energy term $T$ and the potential
energy term $V(0,r)$ on the initial hypersurface are given by

\be
T \equiv {\dot {R}}^2, \; \; V(0,r) \equiv - \frac{F(r)}{r}
\ee

Then from equation (3) we can write,

\be
T = - V(r) + f(r)
\ee

Using the above two expressions we can write,

\be
s(r) = \frac{f(r)}{V(0,r)} = \frac{1}{1 - T/f(r)}
\ee

The function $s(r)$ takes values between 0 and 1. Since $s'(r) \geq 0$ (it is
zero for homogeneous and marginally bound case), it is clear from the above
equation that as we move away from the center to the boundary of the star,
contribution of the kinetic energy term to total energy decreases.
Physically this implies that we do not give additional velocity to the outer
layers of the matter in the cloud, 
in comparison to the inner layers, to avoid shell crossings.
\bigskip

\centerline{\bf V. Naked Singularities and Black Holes}
\bigskip

The singularity theorems prove only the existence, and do not give
any information on
the nature and behaviour of the singularities they predict. 
This leaves an important gap between
the theoretical existence and physical presence of the
singularities of cosmology and gravitational collapse. 
As mentioned earlier, one of the main differences lie in the issue
related to genericity. As we discussed, in dust models the most general 
solution to
Einstein equations contains two free functions, which are functions of the
comoving radial coordinate $r$. If the final outcome of gravitational 
collapse is specified in
terms of both the functions (in a general form), the solution can be called as
functionally generic (see e.g. \cite {Ori} also), 
at least within the context of the dust models we are discussing.

We have discussed in Section III the conditions on the initial data
for the formation of a naked singularity or a black hole. 
As was pointed out there, the behaviour of the first point $r =0$ 
of the singularity curve depends on the initial data,
and all other points on the curve are covered by the apparent horizon
independently of the initial conditions.

Consider now a generic density profile, that is,

\be
\rho(r) = \sum_{n=0}^{\infty} \rho_n r^n
\ee

There are physical reasons to avoid the $\rho_1$ term from the above expansion,
namely that if this term is nonzero, there will be a cusp in the density
at the center of the cloud. However, for the sake of generality, we shall
not assume this term to be necessarily zero. 
Corresponding expression for $F(r)$, and the expression for $f(r)$ 
are then of the following form,

\be
F(r) = \sum_{n=0}^{\infty}F_n r^{n+3}, \: \: f(r) = \sum_{n=2}^{\infty}f_n
r^{n}.
\ee

To analyse the central singularity  $r=0$, we now evaluate the previously
defined quantities $\eta(r)$ , $\beta(r)$ and $p(r)$ for these density
and velocity profiles, which are given by

\be
\eta(r) = 3 + \frac{r F_1}{F_0} + r^2 [\frac{2 F_2}{F_0} -
\frac{{F_1}^2}{{F_0}^2}] + r^3 [\frac{{F_1}^3}{{F_0}^3} - \frac{3 F_1
F_2}{{F_0}^2} + \frac{3 F_3}{F_0}] + O[r]^4
\ee

\be
\beta(r) = 2 + \frac{r f_3}{f_2} + r^2 [\frac{2 f_4}{f_2} -
\frac{{f_3}^2}{{f_2}^2}] + r^3 [ \frac{3 f_5}{f_2} - \frac{3 f_3
f_4}{{f_2}^2} + \frac{{f_3}^3}{{f_2}^3}] + O[r]^4
\ee

and

$$
p(r) \; = \; \frac{f_2}{F_0} + r  \frac{f_2}{F_0} [\frac{f_3}{f_2} -
\frac{F_1}{F_0}]  + r^2 \frac{f_2}{F_0} [\frac{f_4}{f_2} - \frac{f_3
F_1}{f_2 F_0} + \frac{{F_1}^2}{{F_0}^2} - \frac{F_2}{F_0}] ~~~~~~~~~~~~~~~
$$

\be
~~~~~~~~\; + \; \; r^3 \frac{f_2}{F_0} [\frac{f_5}{f_2} - \frac{f_4 F_1}{f_2 
F_0} + \frac{f_3 {F_1}^2}{f_2 {F_0}^2} - \frac{f_3 F_2}{f_2  F_0} -
\frac{{F_1}^3}{{F_0}^3} + \frac{2 F_1 F_2}{{F_0}^2} - \frac{F_3}{F_0}] +
O[r]^4
\ee

respectively. Now consider the expression for $\Theta(r)$

\be
\Theta(r) = \frac{1}{r^{3(\alpha-1)/2}}[\Psi_0 + \Psi_1 r + \Psi_2 r^2 +
\Psi_3 r^3] + \frac{O[r]^4}{r^{3(\alpha-1)/2}} 
\ee

where the expressions for various coefficients of $\Psi$ are the following,

\be
\Psi_0 = 0
\ee

\be
\Psi_1 = [\frac{1}{(1+f_2/F_0)^{1/2}}(\frac{f_3}{f_2} - \frac{F_1}{F_0})
+ (\frac{F_1}{F_0} - \frac{3f_3}{f_2})(\frac{sin^{-1}
{\sqrt{-f_2/F_0}}}{(-f_2/F_0)^{3/2}} + \frac{\sqrt{1+f_2/F_0}}{(f_2/F_0)})] 
\ee

$$
\Psi_2 = [\frac{1}{(1+f_2/F_0)^{1/2}}(-\frac{{f_3}^2}{f_2 F_0} + \frac{f_3
F_1}{{F_0}^2} + \frac{2 f_4}{f_2} - \frac{{f_3}^2}{{f_2}^2} - \frac{2
F_2}{F_0}) + (\frac{2 F_2}{F_0} - \frac{{F_1}^2}{{F_0}^2}~~~~~~~~~~~
$$

$$
- \frac{3 f_4}{f_2} + \frac{3{f_3}^2}{2{f_2}^2}) \times
(\frac{sin^{-1}{\sqrt{-f_2/F_0}}}{(-f_2/F_0)^{3/2}} +
\frac{\sqrt{1+f_2/F_0}}{(f_2/F_0)}) (\frac{3}{2}\frac{sin^{-1}
{\sqrt{-f_2/F_0}}}{(-f_2/F_0)^{1/2}}
$$

\be
~~~~~~~ - \frac{3+f_2/F_0}{2 ({1+f_2/F_0})^{1/2}})
(\frac{f_3 F_0 - f_2 F_1}{{f_2}^2})(\frac{F_1}{F_0} -
\frac{3 f_3}{f_2})]
\ee

$$
\Psi_3 \; = \; [\frac{1}{(1+f_2/F_0)^{1/2}}(-\frac{5 f_3 f_4}{f_2 F_0} +
\frac{{f_3}^3}{{f_2}^2 F_0} + \frac{2 f_3 F_2}{{F_0}^2} -
\frac{3 f_3 {F_1}^2}{2 {F_0}^3} + \frac{F_1{f_3}^2}{2 f_2 {F_0}^2}~~~~~~~~~~~ 
$$ 

$$
+ \frac{3  {f_3}^3}{8 f_2{F_0}^2}  + \frac{F_1 F_4}{2
{F_0}^2}  - \frac{F_1 {f_3}^2}{8 {F_0}^3} + \frac{3 f_5}{f_2} - \frac{3
F_3}{F_0}  - \frac{3 f_3 f_4}{{f_2}^2} + \frac{3 F_1 F_2}{{F_0}^2} +
\frac{{f_3}^3}{{f_2}^3} 
$$

$$
- \frac{{F_1}^3}{{F_0}^3}) 
+ [\frac{5}{4} (\frac{3 sin^{-1}{\sqrt{-f_2/F_0}}}{(-f_2/F_0)^{3/2}}
- \frac{(f_2/F_0)^2 + 4 {f_2/F_0} + 3}{f_2/F_0 (1+
f_2/F_0)^{3/2}})(\frac{f_3}{f_2} - \frac{F_1}{F_0})^2 
$$

$$
+ (\frac{3 sin^{-1} {\sqrt{-f_2/F_0}}}{(-f_2/F_0)^{1/2}} -
\frac{(3+f_2/F_0)}{({1+f_2/F_0})^{1/2}})(\frac{-F_2}{f_2} +
\frac{{F_1}^2}{F_0 f_2} - \frac{F_1 f_3}{{f_2}^2} + \frac{F_0
f_4}{{f_2}^2})] 
$$

$$
\times [\frac{F_1}{F_0} - \frac{3 f_3}{2
f_2}] +  [(\frac{3}{2}\frac{sin^{-1} {\sqrt{-f_2/F_0}}}{(-f_2/F_0)^{1/2}}
- \frac{3+f_2/F_0}{2 ({1+f_2/F_0})^{1/2}})(\frac{f_3 F_0 - f_2
F_1}{{f_2}^2}) \times 
$$

$$
(\frac{2 F_2}{F_0} -
\frac{{F_1}^2}{{F_0}^2} - \frac{3 f_4}{f_2} + \frac{3{f_3}^2}{2{f_2}^2})] + 
(\frac{sin^{-1}{\sqrt{-f_2/F_0}}}{(-f_2/F_0)^{3/2}} +
\frac{\sqrt{1+f_2/F_0}}{(f_2/F_0)}) \times
$$

\be
~~~~~~~~~~~~~~~(\frac{{F_1}^3}{{F_0}^3} - \frac{3 F_1 F_2}{{F_0}^2} + 
\frac{3 F_3}{F_0} -
\frac{9 f_5}{2 f_2} + \frac{9 f_3 f_4}{2{f_2}^2} - \frac{3{f_3}^3}{2{f_2}^3})] 
\ee

These coefficients completely specify the form of $\Theta$, and
$\alpha$ is uniquely determined by the condition that, in the
limit $r \rightarrow 0$, $\Theta$ takes a nonzero finite value.
The existence of naked singularities (black holes) is determined
by the existence (absence) of real positive roots to the equation
$V(X) = 0$. We can write equation (20) as (using the fact that
$f_0 = 0$)

\be
V(X) = (1 - \sqrt{\frac{\Lambda_0}{X}})\frac{H(X,0)}{\alpha} - X = 0
\ee

where the quantity 
$\Lambda_0$ is specified by equation (31), and $H(X,0)$ can be obtained
using equation (14) as

\be
H(X,0) = X + \frac{\Theta_0}{X^{1/2}}
\ee

The equation $V(X)=0$, for which we need to investigate the existence
of real positive roots, then can be reduced to a quartic of the form

\be
(\alpha - 1) x^4 + \sqrt{\Lambda_0} x^3 -\Theta_0 x + \sqrt{\Lambda_0}
\Theta_0 = 0
\ee

where $ x = \sqrt{X}$. The roots, positive or otherwise, of this
equation can be completely specified in terms of the functions
$\Theta_0$ and $\Lambda_0$ at the initial hypersurface. To
analyse the existence of real positive roots in the above
equation, consider a general quartic of the form \cite {Burn}

\be
a x^4 + 4b x^3 + 6c x^2 + 4d x + e = 0
\ee

with the following definitions, $Q = ac - b^2$, $I = ae - 4bd +
3c^2$, $J = ace + 4bcd - ad^2 - eb^2 - c^3$ and $\triangle = I^3 -
27J^2$. The conditions for the existence of real
roots are then, $\triangle < 0$ implies the existence of two real and two
imaginary roots and $\triangle > 0$ implies that all roots are
imaginary unless $Q < 0$ and $(a^2I - 12Q^2) < 0$ (in which case
all roots are real). For the quartic under consideration, we have

\be
\triangle = (\alpha - \frac{3}{4})^3 {\Lambda_0}^2 {\Theta_0}^3
- \frac{27}{(16)^2}[{\Lambda_0}^{3/2} + (\alpha - 1){\Theta_0}^2]^2
\ee

The above analysis only checks for
roots to be real or imaginary; positivity or otherwise of the roots has to be
checked separately for different cases. As explained in Section III, for
$\alpha > 3$ we only have black holes since we cannot have outgoing
geodesics meeting the singularity in the past. Thus, for $\alpha \leq
3$ we can have real positive roots satisfying equation (50).
Consider first the case $\alpha < 3$. From equation (31) we then have
$\Lambda_0 = 0$, and the expression for $\triangle$ simplifies to

\be
\triangle = - \frac{27}{(16)^2} (\alpha - 1)^2 {\Theta_0}^4
\ee

Since both $\alpha$ and $\Theta_0$ take only real values, it follows
that $\triangle < 0 $ for all values of $\Theta_0 $, 
which ensures the existence of real roots. The roots equation $V(X)=0$
gets simplified for the case $\alpha < 3$ as below,

\be
x^{3} = \frac{\Theta_0}{(\alpha - 1)}, \; \; i.e. \; \; X =
[\frac{{\Theta_0}}{(\alpha - 1)}]^{2/3}
\ee

Clearly, for the range $1 < \alpha < 3$, we always have real positive root(s)
to the above equation, ensuring the nakedness of the singularity. 
However, the naked singularities in this case are not strong in the
sense described earlier, since $\Lambda_0 = 0$.

The criterion for the existence of strong curvature singularity is
that $\alpha \geq 3$, and $\alpha = 3$ is the most interesting
case since $\alpha > 3$  give rise to black holes. To analyze this
case, using (50), the roots equation in this case is given as

\be
2 x^4 + \sqrt{F_0} x^3 - \Theta_0 x + \sqrt{F_0} \Theta_0 = 0
\ee

comparing various coefficients with the expression for a general
quartic (51), we have,

\be
a = 2, \; \; b = \frac{\sqrt{F_0}}{4}, \; \; c = 0, \; \; d = -
\frac{\Theta_0}{4}, \; \; e = \sqrt{F_0} \Theta_0
\ee

and we have,

\be
\triangle = - \frac{27}{(16)^2} {\Theta_0}^2 (4 {\Theta_0}^2 -
104 {F_0}^{3/2} \Theta_0 + {F_0}^3)
\ee

If the expression in parenthesis on the right side is greater
than zero, we necessarily have two real roots to the above
equation. To analyse the expression in parenthesis, consider

\be
g(\Theta_0, F_0) = 4 {\Theta_0}^2 - 104 {F_0}^{3/2} \Theta_0 + {F_0}^3
\ee

This equation admits roots $\Theta_1(= \frac{{F_0}^{3/2}}{2}[26 -
5\sqrt{27}])$ and $\Theta_2( = \frac{{F_0}^{3/2}}{2}[26 + 5\sqrt{27}])$,
and both are positive. We can easily see that the
function takes negative values 
for the range $\Theta_1 < \Theta_0 < \Theta_2$, and
is positive outside this range. 
Hence the range $\Theta_0 < \Theta_1$ and $\Theta_0
> \Theta_2$ admits atleast two real roots. 
If $\triangle > 0$, we can have all the roots real if $Q
< 0$ and the quantity $a^2 I - 12 Q^2 < 0$. The first condition is always
satisfied, since $Q = - F_0/16$ (where $F_0 > 0$), and the second condition is
satisfied if $3\sqrt{F_0}(3 \Theta_0 - \frac{{F_0}^{3/2}}{64}) < 0$, i.e.
$\Theta_0 < \frac{{F_0}^{3/2}}{192} (= \Theta_3, say)$. Since $\Theta_3 <
\Theta_1 < \Theta_2$ we cannot have real roots for the range $\Theta_1 <
\Theta_0 < \Theta_2$. To analyse the positivity of given real roots
consider equation (55) which can be written as 

\be
\Theta_0 = x^3 (\frac{2 x + \sqrt{F_0}}{x - \sqrt{F_0}})
\ee

The Fig. 1 explicitly shows the positivity of roots for
equation (55).

\begin{center}
\leavevmode\epsfysize=4in\epsfxsize=4in\epsfbox{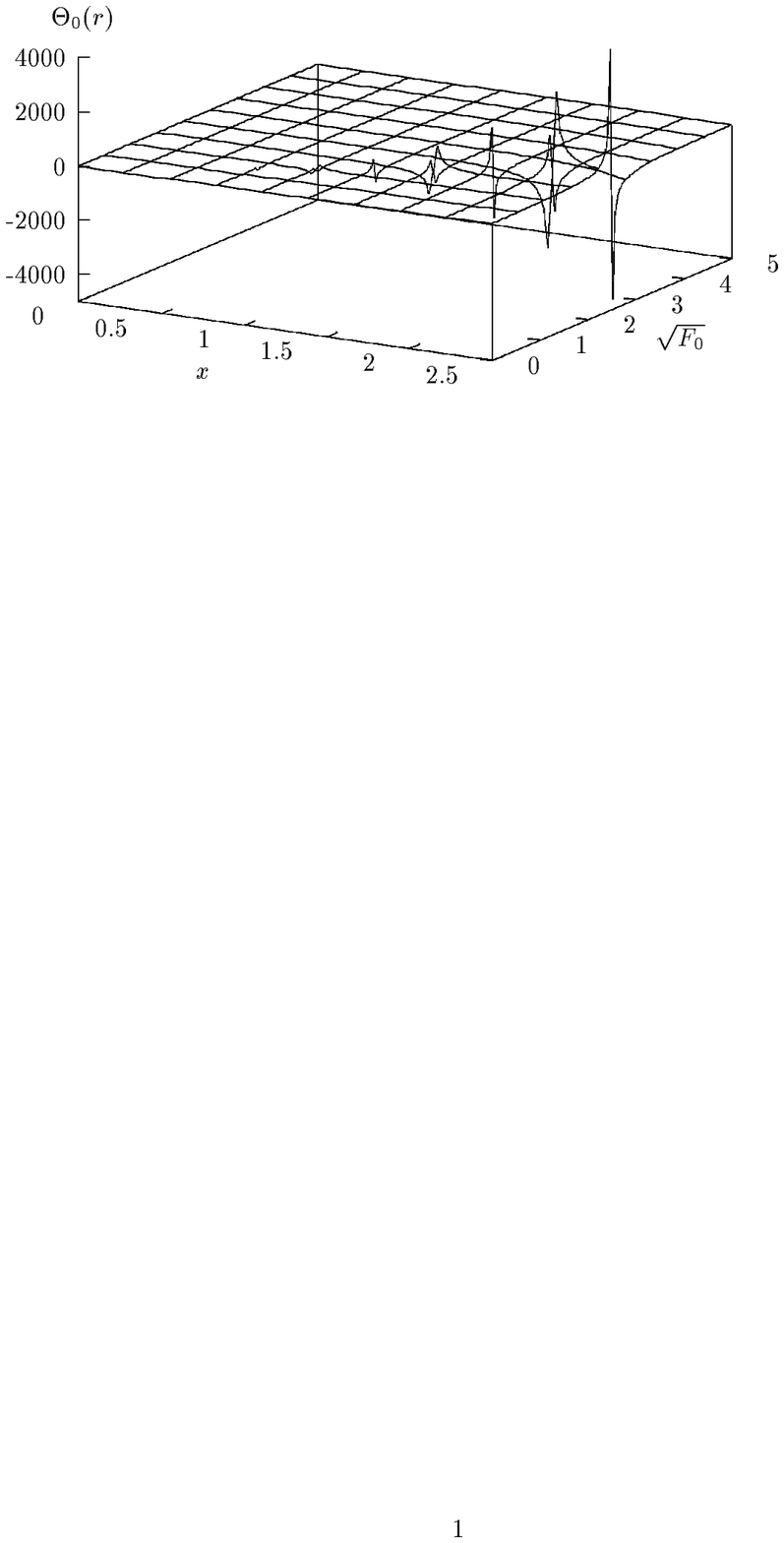}
\end{center}

\noindent {\small Figure 1: The graph showing existence of real positive
roots of equation (55). There are no roots of the form $x = \sqrt{F_0}$
since this implies $F_0 = 0$ which is not true. At all other points on the
hypersurface $x$ admits positive values with non-zero $\Theta_0$ ($x \ne
0$).}

It is easy to analyse from equation (15) that
the regions, in the above figure, where $\Theta_0 $ admits negative values 
necessarily give rise to shell crossings near singularity and the region 
admitting positive values of $\Theta_0$ is free from shell crossings for 
the given initial data. For a further discussion on shell-crossing
singularities, and considerations involving more general class of functions
than considered here, we refer to \cite{Dwivedi96}.

To see more explicitly as to 
how the existence of real positive roots is related
to the data specified at the initial surface, (depending on the choice 
of $F(r)$ and $f(r)$), we now analyse 
some important cases.  
The parameter $\alpha$  will decide the strength of  
the final singularity and we
will be mainly concerned here for the case $\alpha = 3$, i.e. strong
curvature singularities (naked or covered). The value of $\alpha$ can be
uniquely fixed by the limiting value of $\Theta (r)$, which is specified
in terms of various coefficients in the expansion of $\Psi (r)$.
In the expansion of $\Psi(r)$, if the first two terms
$\Psi_1$ and $\Psi_2$ vanish ($\Psi_0 \equiv 0$), and $\Psi_3$ or a higher
term is the first
nonvanishing term,  then necessarily it is a strong
curvature singularity.

The two important specific questions which need to be 
addressed in this connection are, 
(i) Given an arbitrary density distribution on the initial surface,
is it possible to keep the first two terms in the expansion of $\Psi(r)$
zero, and later terms non-zero, by means of a suitable choice of a 
velocity profile for the cloud, so that the resulting singularity
is strong curvature type, and (ii) How `generic' is
the occurrence of the strong curvature naked singularities.

To investigate the answer to the first question, we start by 
considering a specific example. Consider a density profile of the form

\be
\rho(r) = \rho_0 + \sum_{n=2}^{n=\infty}\rho_n r^n, \;\; \rho_2\ne0
\ee

where $\rho_0$ is the central density of the cloud. We will assume the density
to be decreasing outwards from the center, which is a reasonable condition
to take for realistic density profiles, and which is also 
a necessary condition to avoid any possible shell-crossing singularities. 
The choice of such a density
distribution has some physical significance. Firstly, as commented earlier, the 
linear term in the density 
profile is chosen to be zero, because a non-vanishing
linear term would mean a cusp in the density at the center with a non-zero
pressure gradient at the center. This may
not be desirable on physical grounds. Next, for physically realistic density
configurations, the $\rho_2$ term is generally believed to be non-zero,
with $\rho_2<0$, for reasons related to the stability of the stars 
(see e.g. \cite {Gough, Antia}).  
The subsequent terms
in the expansion may be vanishing or otherwise, which will not affect our
considerations here, because as we shall see, it is the first non-vanishing
term in the expansion which determines the nature of the singularity. 
Another useful point to note is that for a marginally bound collapse
($f=0$), such a density profile necessarily results into a weak naked
singularity \cite {JTP}. Hence, it is of importance to know whether the
collapse of the same density profile can result into a strong curvature
naked singularity when the collapse is not marginally bound.

For the density profile given above, the corresponding 
expression for $F(r)$ is given by 

\be
F(r) = F_0 r^3 + \sum_{n=2}^{n=\infty}F_n r^{n+3}
\ee

As for the other free function $f(r)$, the appropriate differentiability
for the metric requires it to
be at least a $C^2$ function. Therefore, a general choice for $f(r)$ is,

\be
f(r) = \sum_{n=2}^{n=\infty}f_n r^n
\ee

We now fix  $f(r)$, which specifies the velocity
profile for the cloud, by imposing the following two requirements: (i) the
second term in the expansion of $f(r)$, i.e. $f_3$ is zero and, (ii)
the coefficients $\tilde{F_2}$ and $\tilde{f_4}$ as given by 

\be
\tilde{F_2} = \frac{F_2}{F_0}, \; \; \; \; \tilde{f_4} = \frac{f_4}{f_2}
\ee

be related as

\be
\tilde{F_2} = \tilde{f_4}\frac{[\frac{1}{(1+f_2/F_0)^{1/2}} -
\frac{3}{2}G(-f_2/F_0)]}{[\frac{1}{(1+f_2/F_0)^{1/2}} - G(-f_2/F_0)]}
\ee

Under this situation, we evaluate the expressions for $\eta(r)$ and 
$\beta(r)$, which turn out to be

\be
\eta(r) = 3 + r^2 [2 \tilde{F_2}] + r^3 [\frac{3 F_3}{F_0}] + O[r^4]
\ee

\be
\beta(r) = 2 + r^2 [2 \tilde{f_4}] + r^3 [\frac{3 f_5}{f_2}] + O[r^4]
\ee

respectively. Then, considering the  general expression for $\Theta(r)$,

\be
\Theta(r) = \frac{\Psi(r)}{r^{3(\alpha-1)/2}}
\ee

the various coefficients are evaluated to be as follows

\be
\Psi_0 = 0
\ee

\be
\Psi_1 = 0
\ee

\be
\Psi_2 = 0
\ee

\be
\Psi_3 = 3 (\frac{f_5}{f_2})[\frac{1}{(1+f_2/F_0)^{1/2}} -
\frac{3}{2}G(-f_2/F_0)] 
\ee

Since $\Psi_3$ is the first nonvanishing term in the expansion of $\Psi(r)$,
we have  $\alpha = 3$, which makes it to be a strong curvature singularity.
Also, whenever $\alpha=3$ we know that $\Lambda_0$ is a nonzero quantity, so
from the roots equation $V(X) = 0$ we have, 

\be
(1 - {\sqrt{\frac{\Lambda_0}{X}}}) \frac{H(X,0)}{3} - X = 0 
\ee

where $H(X,0)$ is given by

\be
H(X,0) = X + \frac{\Theta_0}{X^{1/2}}
\ee

Writing again $x= \sqrt X$, this becomes a quartic equation,

\be
2x^4 + {\sqrt {F_0}}x^3 - \Theta_0 x + \Theta_0 {\sqrt {F_0}}
= 0
\ee

Depending on the choice of the values of the initial variables, that is 
$f_5, f_2$ and $F_0$, we
can have the above quartic with or without real positive roots (Fig. 1). Hence
we can have both the possibilities as we would desire, namely the
black holes as well as naked singularities, depending on the choice 
of initial data. As we have already pointed out earlier, in either case
this is a strong curvature singularity.

We have illustrated above how strong curvature
singularities arise as a result of collapse when the coefficient 
$\rho_2$ is the 
first nonvanishing term in the expansion
of the density profile. 
We chose in the above the $\rho_1$ term in the expansion of density
to be identically zero, since it would represent a cusp in the density at the
center of the star and may be objectionable on physical grounds. However,
it is not conclusively established that such density cusps at the 
center are ruled out by astrophysical considerations (see e.g. \cite {Spitzer})
Hence, if we assume that the $\rho_1$ term is non-vanishing and
arises as a perturbation (which can possibly
happen in the turbulent interiors of stars), in that case as well we can show
that strong curvature singularities can arise as the final fate of collapse.
For that purpose, using similar techniques as above, one  again chooses the 
velocity profile suitably, imposing required constraints on $f(r)$. 
Next, when $\rho_3$ is the first
non-vanishing term in the density expansion, 
then it is already known that we have a strong curvature singularity 
for the marginally bound collapse. This case is discussed earlier 
in detail \cite {Singh}, showing the formation of naked
singularities and black holes. In a subsequent paper, the  causal structure
of spacetime has been analysed, using the dynamics of trapped surfaces, for
this particular example \cite {Jhingan} (see Fig. 2)

\begin{center}
\leavevmode\epsfysize=4in\epsfxsize=4in\epsfbox{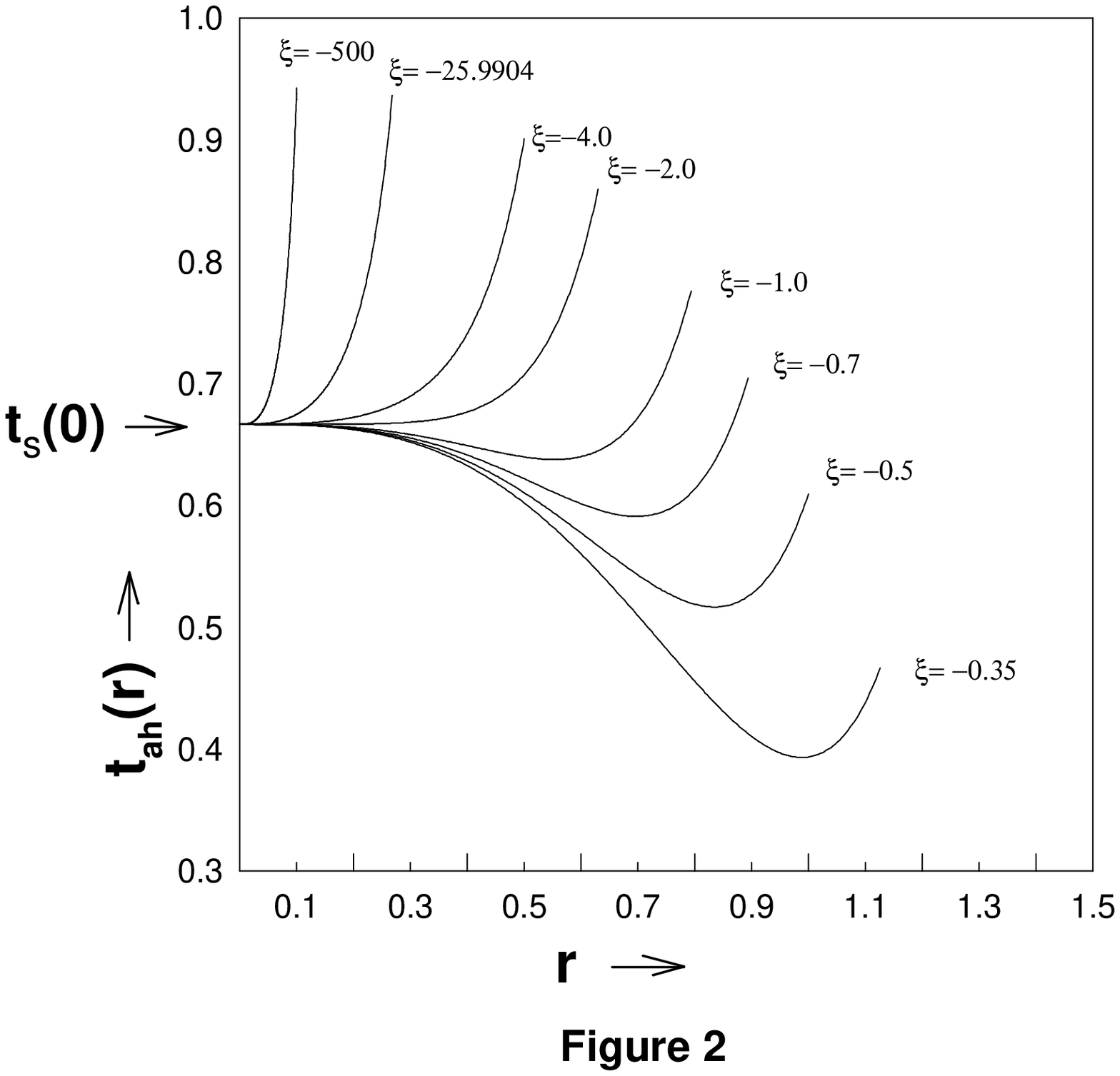}
\end{center}

\noindent {\small Figure 2: A plot of the apparent horizon curves (12) 
for marginally bound case and , $\rho(r) = \rho_{0}\left( 1 -
{r^{3}\over r_{c}^{3}}\right)$, obtained by setting $\rho_0=1$ and varying
$r_c$ (boundary of the cloud). For $\xi<-2$ $\xi (= F_3/{F_0}^{5/2})$ the 
center is the first point to get trapped, whereas for $\xi>-2$ some
surface in the 
interior of the star is the first one to get trapped, and the trapped
region moves both inwards and outwards. But for all other points on the
singularity curve ($r > 0$), $t_s(r) > t_{ah}(r)$, independent of the initial
data.}
 
\bigskip

Interesting situation arises when all the first three terms $ \rho_1,\rho_2$
and $ \rho_3 $ in the 
expansion of density profile are zero, and the first non-vanishing term in the
expansion of density is $\rho_n \; (n>3)$. We know that in the marginally bound
case this situation always corresponds to a black hole \cite {JTP}.  
Now, considering the general non-marginally bound case, the density profile 
is of the form

\be
\rho(r) = \rho_0 + \sum_{n=4}^{n=\infty}\rho_n r^n
\ee

The corresponding expression for $F(r)$ can be written as 

\be
F(r) = F_0 r^3 + \sum_{n=4}^{\infty} F_n r^{n+3}
\ee

Consider then the following  choice of the energy function,
or equivalently the particle velocity profile, as given by

\be
f(r) = f_2 r^2 +\sum_{n=5}^{n=\infty} f_n r^n
\ee

We can see that for this choice of $f(r)$ (i.e. $f_3 =0, \; f_4 =0$)  we will
have $\Psi_0$, $\Psi_1$ and $\Psi_2$ identically zero, and $\Psi_3$ is
nonzero with

\be
\alpha = 3, \;  \; \Psi_3= \frac{3 f_5}{f_2}[\frac{1}{(1 + f_2/F_0)^{1/2}} -
\frac{3}{2}G(- f_2/F_0)]
\ee

The expression for quartic remains the same equation (55), except that
the functional form
of $\Theta_0$ changes now. Hence, we can have again both the possibilities,
namely the black holes and strong curvature naked 
singularities depending on the choice of initial free functions.

After starting the gravitational 
collapse from a regular initial data, it is desirable to
ensure that no shell-crossing singularity forms in the cloud before
formation of central shell-focusing singularity. The necessary and sufficient 
condition to avoid shell-crossings in the cloud is given by equation (11), 
which can again be written as (equation (34)),

$$
q'(r) \leq 0 \; , \; s'(r) \geq  0
$$

The condition $q'(r) \leq 0$ 
implies that the density should decrease or remain constant as we 
move away from the center of the cloud. This condition is satisfied 
as long as the sum
of all non-vanishing derivatives is negative (or zero), keeping the overall 
density
positive (to satisfy the weak energy condition); or if all the coefficients
in the density profile are negative. 
We note that the condition for existence
of strong curvature singularity derived here depends on 
the first non-vanishing term of density profile,
which is taken negative in our case, and the remaining terms would not 
contribute to
the final outcome. Hence, rest of the terms in the density profile can
always be chosen to satisfy this condition by the initial
data $F(r)$ chosen in the analysis. Now consider the second
constraint i.e., $s'(r) \geq 0$. As discussed earlier, as we
go out from the center of the cloud, this condition implies that the
contribution of kinetic energy to the total energy decreases. This 
reduces to the following condition on the initial data,

\be
{F(r)} \leq C {[- r f(r)]}
\ee

Since $F(r)$ and $- r f(r)$ admit only positive values, $C$ here is a 
non-zero positive constant. A lower bound on the values of $C$ easily
follows using the boundary condition at the surface of the star, i.e.

\be
C \geq \frac{M}{-r_c f_c}
\ee

where M is the total mass of the cloud and the subscript $c$ denotes values
measured  at the boundary of the cloud. In equation (80), $M/r_c$ and $f_c$ 
denote respectively the potential energy and the 
total energy of the shell at the boundary of 
the cloud respectively. In general, from equations (3) and (79) we can 
see that we must have $C \geq 1$ always, equality being 
satisfied for shells which are rest. Also, from equation
(80), the equality implies that  boundary of the star is at rest.

Let us consider a specific example to understand the possibility of existence
of regular initial data ($F(r)$, $f(r)$) for which there are no shell-crossing
singularities in the cloud till the first shell-focusing singularity forms 
at the center of the cloud, i.e.

$$
t_s(0) <  t_{sc}(r)|_{r>0} 
$$

At the center of the cloud we have $r = 0, t_s(0) = t_{sc}(0)$.
Consider a density profile of the form (60). Corresponding
expression for mass function is given by (61). For energy function 
$f(r)$ of the form (62) subject to suitable constraints ($f_3 = 0 $) and (64),
we have shown earlier 
that it gives rise to strong curvature naked singularity. 
The condition (79) for avoiding shell-crossing singularities reduces to

$$
F_0 + F_2 r^2 + \cdot \cdot \leq C[- f_2 - f_4 r^2 - f_5 r^3 + \cdot \cdot]
$$

where $C \geq 1$. The above condition will hold in general, everywhere in
a cloud of arbitrary radius, at least for that class of functions, $f(r)$,
where each coefficient in the expansion of $f(r)$ satisfies the above
inequality  with corresponding coefficients in the expansion of $F(r)$.
This can be easily done in this case, since $f_5$ can be given any
finite value (keeping $-1 < f(r) < 0$) which makes $\Psi_3 \neq 0$ in (71), 
and since both $f_5$ and $f_2$ are less then zero therefore $\Psi_3 >
0$, which is consistent with  $\Theta_0 > 0$. Since $C \geq 1$, the argument of
$G(y)$ ($y = - f_2/F_0$) in (64) also satisfies the condition $0 < y \leq 1$.

Now consider another example where the density profile is given by (75).
Here we make a choice of $f(r)$ such that the coefficients $f_3$ and $f_4$ 
are identically zero. Clearly, the 
condition (79) is trivially satisfied as the
corresponding coefficients in the expansion of $F(r)$ are also zero. So we
are free to make a suitable choice of  $|f_5|$, and $|f_2|$ (non-zero) for 
any given $F(r)$ such that the required inequality (80) is satisfied. 
Hence the condition for avoiding shell-crossing singularities 
can always be satisfied in general without affecting the general 
conclusions regarding the generic existence of strong curvature singularities.

This completes our answer to the first question, which is in the affirmative.
That is, for any given density distribution at the initial epoch of time,
we can always keep the coefficient $\Psi_3$ as the first non-vanishing 
term in the expansion of $\Psi(r)$, by means of a suitable initial choice 
of the velocity profile $f(r)$ for the cloud. In other words, given any
generic density profile, we can always choose the rest of the initial data,
that is the particle velocity profile, in such a manner that the 
final fate of the collapse results in a strong curvature singularity which
is either hidden inside a black hole, or naked, communicating with faraway
observers in the spacetime.

Coming to question (ii), the following consideration may provide
some insight into the issues regarding genericity. Consider the functional 
form of the initial data, as we have discussed above.
The functions $F(r)$ and $f(r)$ both are specified in terms of infinitely
many free functions in the form of the coefficients of the expansion
terms such as $F_0, F_1, F_2,...$, and $f_2, f_3,...$ etc.
Hence, we have here a $2 \otimes \infty$ dimensional initial data
space. Now keeping $F(r)$ completely free and fixing finite number of 
coefficients in $f(r)$, as we have done above, to demonstrate the existence
of strong naked singularities still leaves us with a $2 \otimes \infty$ 
dimensional parameter space, and hence the final result in terms of
either a naked singularity or black hole is stable under
a range of perturbations in the newly constrained parameter space.
We would like to contrast this situation with the possible scenario
when we could possibly have only a finite dimensional initial data space
which generates a strong naked singularity, as a subset of an infinite
dimensional initial data space. Then, since the strong curvature singularity
occurs only for a finite number of coefficients in the space of finitely
many coefficients characterizing the complete space of
initial data, we could possibly say that any generic point in the 
complete space will not lie on such a hypersurface of strong singularity, 
and hence the data generating strong singularities must be a set of 
``measure zero" in some sense. Such an argument gives an idea on the
genericity of the occurrence of strong curvature singularities
in gravitational collapse which could be either naked or covered.
\bigskip

\centerline{\bf VI. Discussion and Conclusion}
\bigskip

We have used here collisionless fluid, that is dust with $p=0$ equation 
of state, as the model to analyse the issue of the final fate of gravitational
collapse of a massive cloud, if the collapse started from a
``physically reasonable" initial data in terms of the initial density and
velocity profiles of the cloud. We have imposed only rather general
and physically reasonable conditions such as the matter satisfies the
weak energy condition, and that the collapsing dust avoids shell-crossing
singularities or caustics (this is essentially because our main focus
of interest is the nature of the shell-focusing singularity occurring at 
the center). For the sake of clarity, and to generate maximum possible 
physical insight, we have confined here the attention to the class of Taylor
expandable density and velocity profiles only and considered mainly the
occurrence of strong curvature singularities which are naked or hidden 
inside black holes.

An important limitation of the analysis presented here could be thought 
of as the choice of the dust equation of state. The question as to whether
the introduction of pressure will significantly 
modify the conclusions given here
needs to be looked into carefully. It may be pointed out,
however, that one of the motivations for not considering the general equation 
of state presently is the possibility indicated by recent work
\cite {Dwiv} that the specific properties of matter fields may not turn 
out to be the key factor in deciding the final fate of gravitational collapse.
The indications by the work such as above are that the nature of the central
singularity should essentially depend on the choice of initial data, and 
also there is some kind of a pattern in the behaviour of the central point
($r = 0$), and the other points on the singularity curve, regardless of the
exact form of the matter used \cite {Chris, Josh93, Dwiv, detref}.  
Again, while deciding on the issue of how significant (or insignificant) 
the assumption of dust is, the arguments such as those by Hagedorn \cite {Hag} 
need to be considered that a very soft equation of state is a
good approximation under extreme physical conditions of the advanced state
of collapse. It is possible that even if we include pressures,
the collisions also will give a contribution to the energy-momentum 
tensor, and will accelerate the formation of singularities \cite {Misner,
Teuk}. Hence the collisionless assumption could represent fairly general
features of collapse and deserves serious consideration from the
point of view of possible physical applications. 
From such a perspective, if we assume the formation of singularity
in collapse, as implied by the singularity theorems as well as the physical
considerations on the final fate of collapsing 
massive stars which have exhausted their nuclear fuel, then
we have tried to argue here that ``actual and real" strong curvature
singularities (naked or covered) do appear in a rather ``generic"
way as the end product of gravitational collapse.
%%%%%%%%%%%%%%%%%%%%%%%%%%%%%%%%%%%%%%%%%%%%%%%%%%%%%%%%%%%%%%%%%%%%%%%%%
\begin{flushleft}
{\large Acknowledgements:}
We thank I. H. Dwivedi and T. P. Singh for comments and useful
discussions.
\end{flushleft}
%%%%%%%%%%%%%%%%%%%%%%%%%%%%%%%%%%%%%%%%%%%%%%%%%%%%%%%%%%%%%%%%%%%%%%%%%

%%%%%%%%%%%%%%%%%%%%%%%%%%%%%%%%%%%%%%%%%%%%%%%%%%%%%%%%%%%%%%%%%%%%%%%%
\end{document}